\documentclass[11pt,twoside]{article}
\usepackage{asp2010}
\usepackage{epstopdf}
\usepackage{graphicx}
\usepackage{epsf}

\newcommand{\ldl}{$\lambda/{\Delta}{\lambda}$}

\newcommand{\teff}{T$_{eff}$}

\newcommand{\meth}{CH$_4$}
\newcommand{\wat}{H$_2$O}

\newcommand{\kms}{km~s$^{-1}$}

\newcommand{\mjup}{M$_{Jup}$}

\resetcounters

\begin{document}

\title{Cool Star Science with the FIRE Spectrograph}
\author{
Adam J.\ Burgasser$^{1,2}$,
Robert A.\ Simcoe$^{2}$,
John J.\ Bochanski$^{2,3}$,
Carl Melis$^{1,4}$,
Craig McMurtry$^{5}$,
Judy Pipher$^{5}$,
William Forrest$^{5}$,
Michael C.\ Cushing$^{6}$,
Dagny L.\ Looper$^{7}$,
and Subhanjoy Mohanty$^{8}$
\affil{$^1$Center for Astrophysics and Space Sciences, UC San Diego, 9500 Gilman Drive, La Jolla, CA 92093, USA (aburgasser@ucsd.edu)}
\affil{$^2$Kavli Institute for Astrophysics and Space Research, Massachusetts Institute of Technology, 77 Massachusetts Avenue, Cambridge, MA 02139, USA}
\affil{$^3$Dept.\ Astronomy \& Astrophysics, Pennsylvania State University, 525 Davey Laboratory, University Park, PA 16802, USA}
\affil{$^4$CASS Postdoctoral Fellow and NSF Astronomy and Astrophysics Postdoctoral Fellow}
\affil{$^5$Dept.\ of Physics and Astronomy, University of Rochester, Rochester, NY 14627, USA}
\affil{$^6$NASA Jet Propulsion Laboratory, California Institute of Technology, Pasadena, CA 91109, USA}
\affil{$^7$Institute for Astronomy, University of HawaiÕi, 2680 Woodlawn Dr., Honolulu, HI 96822, USA}
\affil{$^8$Imperial College London, 1010 Blackett Lab., Prince Consort Road, London SW7 2AZ, UK}
}

\begin{abstract}
The Folded-port InfraRed Echellette (FIRE) has recently been commissioned on the Magellan 6.5m Baade Telescope.  This single object, near-infrared spectrometer simultaneously covers the 0.85-2.45~$\mu$m window in both cross-dispersed ({\ldl} $\approx$ 6000) or prism-dispersed ({\ldl} $\approx$ 250-350) modes.  FIRE's compact configuration, high transmission optics and high quantum efficiency detector provides considerable sensitivity in the near-infrared, making it an ideal instrument for studies of cool stars and brown dwarfs. ÊHere we present some of the first cool star science results with FIRE based on commissioning and science verification observations, including evidence of clouds in a planetary-mass brown dwarf, accretion and jet emission in the low-mass T Tauri star TWA~30B, radial velocities of T-type brown dwarfs, and near-infrared detection of a debris disk associated with the DAZ white dwarf GALEX 1931+01.
\end{abstract}

\section{The FIRE Spectrograph}

\begin{figure}[!ht]
\plottwo{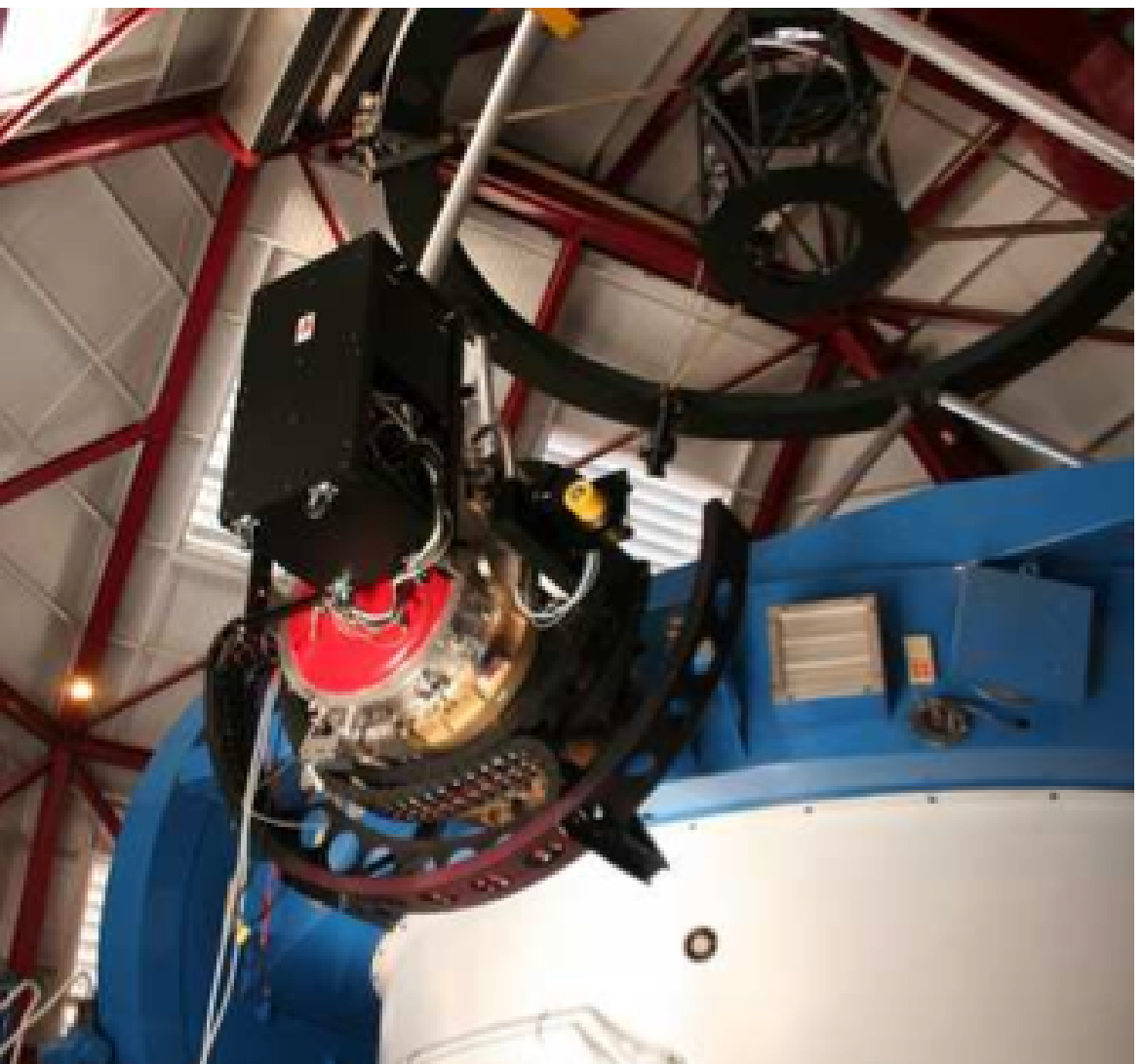}{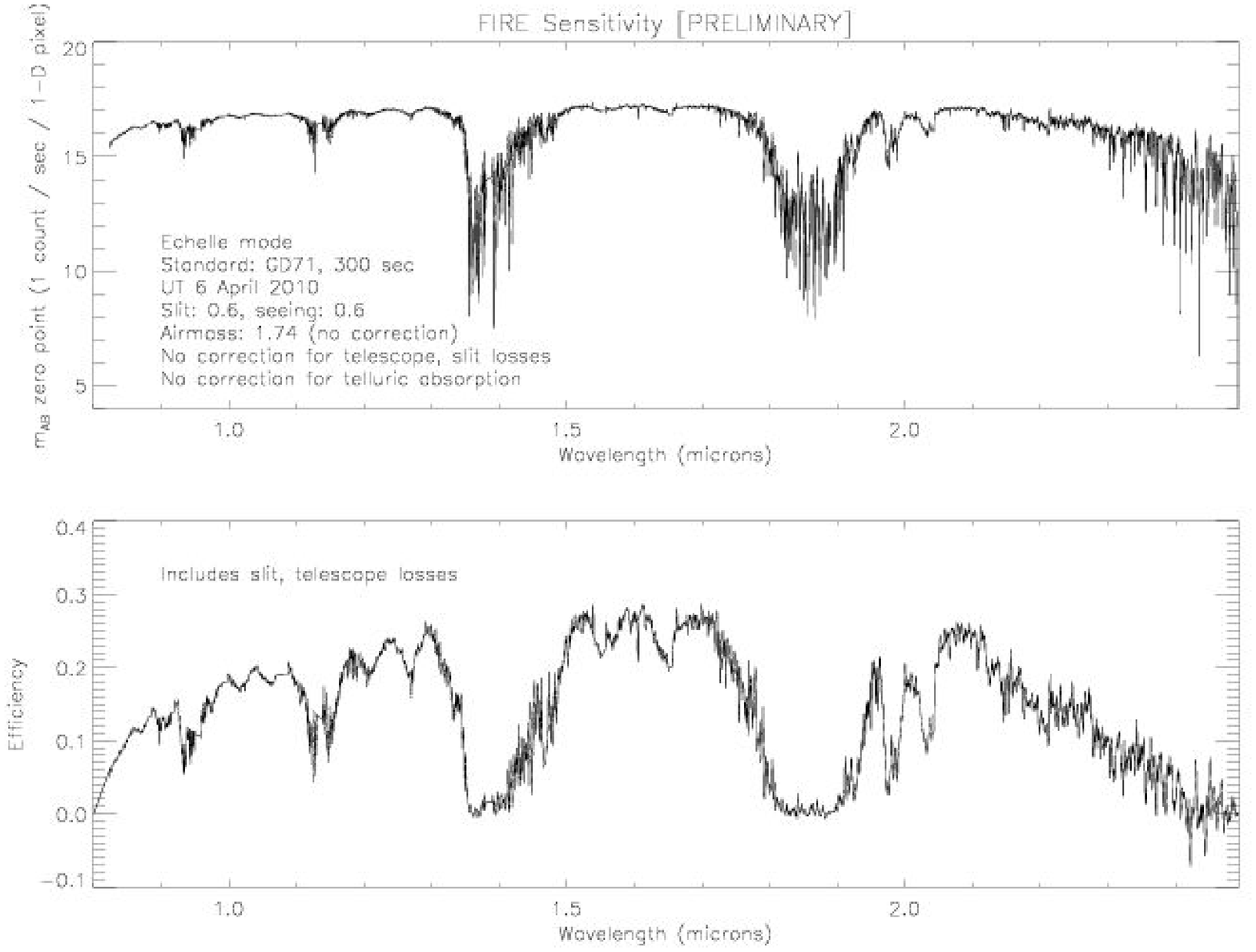}
\caption{(Left) FIRE mounted on the Nasmyth auxiliary port of the Baade telescope.  The cryostat (circular structure with red cap and gold rim) is roughly 1~m in diameter.  The mounted black cabinet holds electronics, while the lower rim carries cables and allows for rotation of the entire instrument.  (Right): Preliminary measurement of FIRE's zeropoint sensitivity (1~count/second/pixel) in its cross-dispersed mode, based on observations of spectrophotometric standard GD~71. The prism-dispersed mode zeropoint is roughly 3-4~mag fainter, although higher background limits sensitivity at long exposures.
\label{fig_fire}}
\end{figure}

In March 2010, the Folded-port InfraRed Echellette (FIRE) was successfully commissioned on the Magellan 6.5m Baade telescope, following three years of design, development and construction at MIT and the University of Rochester.   FIRE is a single-object, near-infrared spectrometer designed to cover the entire 0.85-2.45~$\micron$ window in a single exposure, at a resolution optimizing the balance between telluric OH absorption avoidance and sensitivity for faint near-infrared sources.  Its primary cross-dispersed mode delivers moderate-resolution spectra, {\ldl} $\approx$ 6000 (for the 0$\farcs$6 or 4-pixel wide slit), across 21 orders, simultaneously imaged onto a single HAWAII-2RG detector. A prism-dispersed mode (echelle grating replaced with a mirror) delivers low-resolution spectra, {\ldl} $\approx$ 250-350 (varying across the detector), at higher throughput. Preliminary zeropoints (1~count/second/pixel) for these two modes are AB $\approx$ 17 and AB $\approx$ 20-21, respectively, based on observations of the near-infrared spectrophotometric flux standard GD 71 (Figure~\ref{fig_fire}).  These correspond to 1~hr limiting magnitudes for signal-to-noise S/N $\approx$ 5 of $\sim$21 and $\sim$23-24, respectively (note the higher background penalty at low resolution).  A second near-infrared camera images the entrance slit through an MKO $J$-band filter for source acquisition and manual guiding, and can easily acquire sources with $J$ = 19--20 (AB) in 20--30~s exposures. An external calibration unit is mounted to the spectrograph, with a fold mirror used to direct quartz flat field and ThAr lamp light into the instrument for calibration of the cross-dispersed mode (calibration of the prism-dispersed mode is done using lamps reflected off of the Baade dome spot). 
The entire instrument is compact ($\sim$1~m across) and mounted on one of three f/11 auxiliary (folded) ports along the mirror support structure (Figure~\ref{fig_fire}).  FIRE is continuously available and directed on-sky using a tertiary fold mirror in the telescope, allowing for target-of-opportunity observations and integration with other Baade instruments (e.g., the forthcoming FourStar imager; \citealt{2008SPIE.7014E..95P}).
A data reduction pipeline based on MASE \citep{2009PASP..121.1409B} is nearing completion and is expected to be delivered in early 2011.  
%This pipeline generates a 2D wavelength map for each pixel, a 2D model of the sky though b-spline fitting \citep{2003PASP..115..688K}, performs iterative optimal extraction, stitches the orders, and computes relative flux calibration and telluric correction using the {\em xtellcor} module of SpeXtool \citep{2003PASP..115..389V, 2004PASP..116..362C}.
Further details on FIRE's design and construction can be found in \citet{2008SPIE.7014E..27S, 2010SPIE.7735E..38S} and on the instrument webpage, {\em http://www.firespectrograph.org}.

%FIRE's design was aimed at facilitating study of rare, near-infrared sources - the lowest-temperature stars and brown dwarfs and the highest-redshift quasars.  As verified in our commissioning and science verification runs, it is also well-suited for observation of Kuiper Belt objects, young stars with accretion signatures, eclipsing binaries, supernovae and low redshift galaxies.  Below, we describe some of the preliminary cool star science achieved from FIRE's initial commissioning observations.

\section{Early Cool Star Science with FIRE}

\subsection{Clouds in Planetary-Mass Brown Dwarfs}
FIRE's prism-dispersed mode allows single-order measurement of the near-infrared spectral energy distributions of faint brown dwarfs, including L- and T-type dwarfs, with resolution sufficient to study molecular band and atomic line features, and broadband spectral structure.  The latter is relevant to the study of mineral condensate clouds present in the cool photospheres of these sources (e.g., \citealt{1989ApJ...338..314L,1996A&A...305L...1T, 2001A&A...376..194H}).  Condensate grains reveal themselves in the near-infrared by modulating broadband spectral color and {\wat} band absorption.  The influence of clouds on emergent spectra is prominent for the warmer L-type dwarfs (e.g., \citealt{1998A&A...337..403B, 2002ApJ...568..335M}), but clouds are presumed to have sunk below the photospheres
of T dwarfs cooler than $\sim$1000~K (\citealt{2001ApJ...556..872A, 2006ApJ...640.1063B}).  Our early results with FIRE suggest that young T dwarfs may in fact have significant cloud opacity at their photospheres.

\begin{figure}[!ht]
\plotone{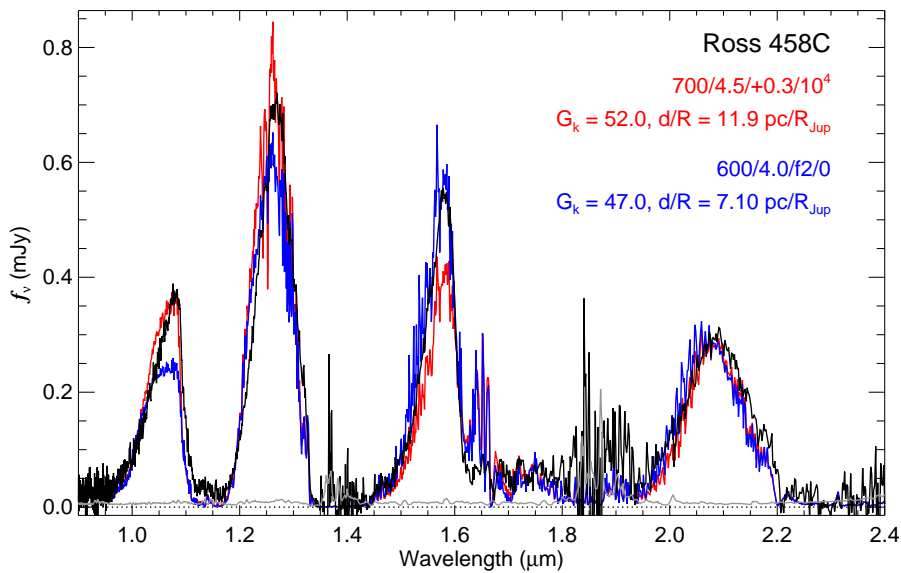}
\caption{FIRE spectroscopy of the T8 dwarf Ross~458C (black line) compared to best-fit models 
from \citet{2008ApJ...689.1327S} with (blue line) and without (red line) cloud opacity.  The cloudy model more accurately reproduces the relative $J-$ to $H-$band peak fluxes, the result of increased opacity at the shorter wavelength peak.  The inferred effective temperature from the model fits, {\teff} = 635$^{+25}_{-35}$~K, is consistent with estimates based on the luminosity and age of the source.
\label{fig_ross}}
\end{figure}

During our March 2010 commissioning run, we obtained prism spectra of the brown dwarf candidate Ross~458C \citep{2010A&A...515A..92S, 2010MNRAS.405.1140G}, a widely-separated (102$\arcsec$), common proper motion companion to the nearby M0.5Ve + M7 Ross~458 binary.  Due to the high throughput of FIRE's prism mode, exposures were limited to 150~s to avoid saturating telluric OH lines in the $H$-band region\footnote{Subsequent observations suggest that prism exposures should be limited to 120~s in typical observing conditions to avoid nonlinearity in the OH region.}, although only four exposures (10~min total integration) were needed to obtain S/N $\approx$ 50-80 in the $YJHK$ flux peaks for this $J = 16.7$ source (Figure~\ref{fig_ross}).  The observations identify Ross~458C as a T8 dwarf, and provide sufficient resolution to measure its weak 1.25~$\micron$ K~I doublet lines.  The inferred classification and known distance of Ross~458C permit an estimate of its bolometric luminosity, while the 150--800~Myr age of the system as inferred from the active, rapidly-rotating primary further indicate a low effective temperature ({\teff} = 650$\pm$25~K) and low mass (M = 6--11~{\mjup}) for this source.  With a projected separation of 1100~AU, Ross~458C is the most widely-separated planetary-mass companion to a nearby star identified to date \citep{2010arXiv1009.5722B}.

The FIRE spectrum of Ross~458C was compared to atmospheric models from \citet{2008ApJ...689.1327S}, with separate comparisons made to models with and without cloud opacity.  We found that cloudy models provide significantly better fits to the spectrum than the cloud-free models, more accurately reproducing the relative $J$- and $H$-band peak fluxes.  This is the first spectroscopic evidence of clouds in the photosphere of a low-temperature brown dwarf and in a planetary-mass object.  However, Ross~458C may be a special case, as its youth and supersolar metallicity (also inferred from the properties of its primary) may result in higher clouds with larger grains (more opacity) and possibly more condensate material.  Similar arguments have been proposed to explain the apparently thick clouds of
young, low-gravity L dwarfs \citep{2006ApJ...651.1166M, 2008ApJ...678.1372C}.  

\subsection{Accretion and Jet Emission in a Low-mass T Tauri Star}

\begin{figure}[!ht]
\plottwo{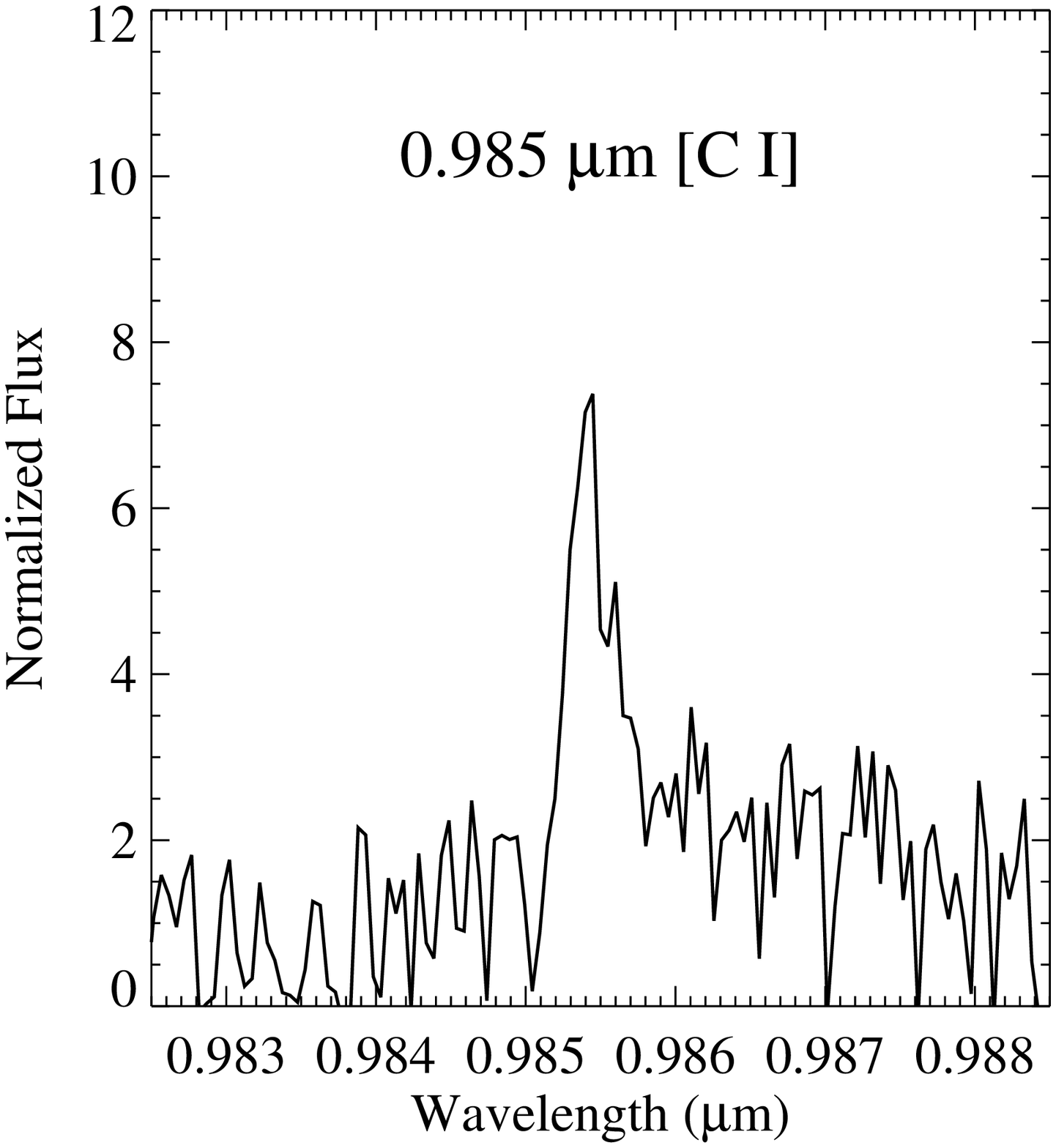}{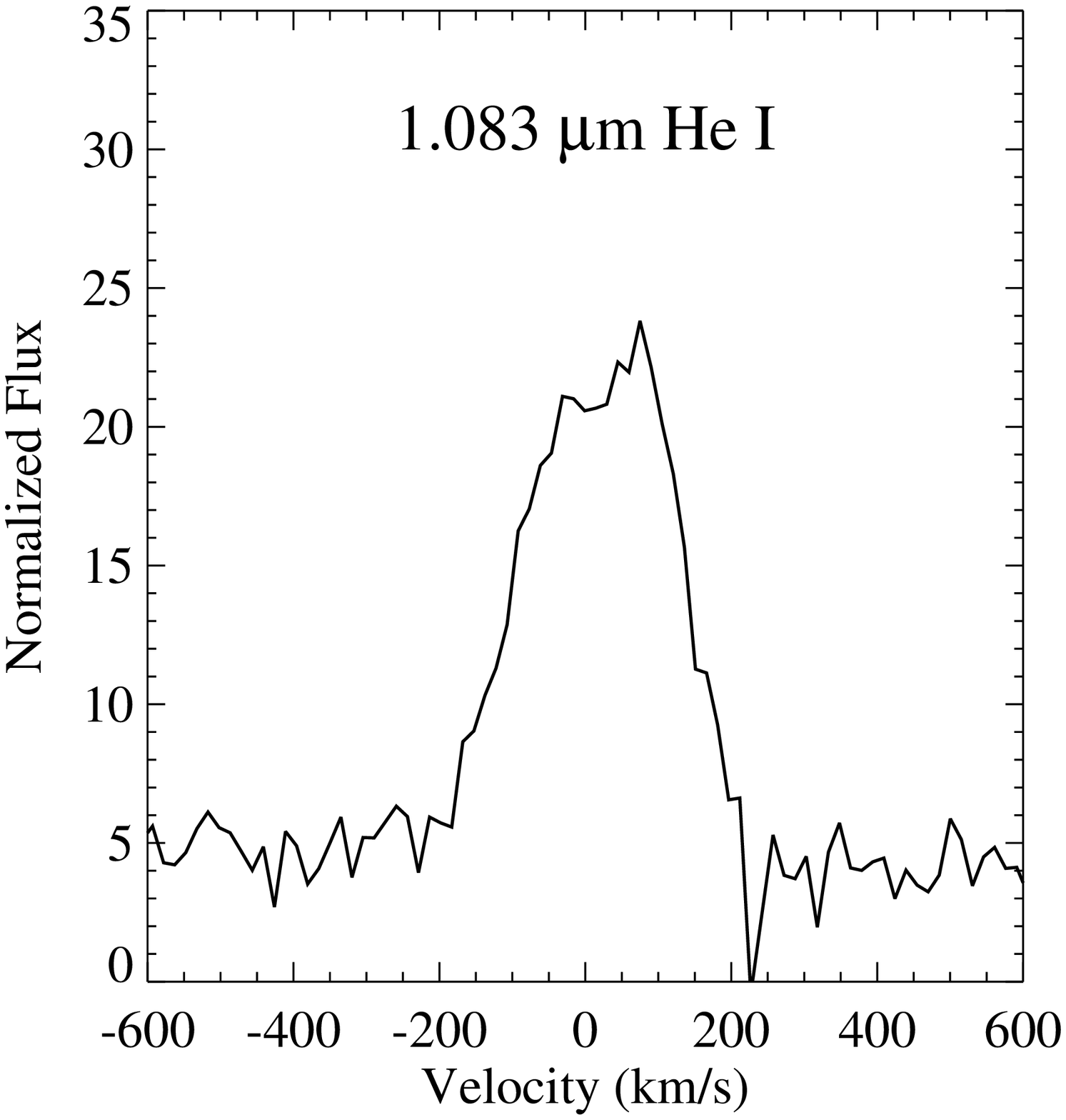}
\caption{Accretion signatures in the spectrum of the low-mass TW Hydrae object TWA~30B: 0.985~$\micron$ [C~I] (left) and 1.083~$\micron$ He~I (right).  The [C~I] line traces collisionally excited gas in the inner regions of the disk, and compared to mid-infrared [C~I] emission can provide a measure of the size of the gaseous disk. The He~I line is heavily broadened, evidence of ongoing accretion from an edge-on disk.
\label{fig_twa30}}
\end{figure}

Actively accreting low-mass stars exhibit a broad range of emission features arising from inflow onto the central star, jets and disk/star interactions via magnetic fields (e.g., \citealt{2003ApJ...589..397K, 2007ApJ...659L..45W}).  Many of these emission diagnostics are found at near-infrared wavelengths, including line and molecular band emission (e.g., \citealt{2003ApJ...589..931N, 2008ApJ...687.1117F}).   We used the moderate-resolution mode of FIRE to search for these diagnostics in the occluded low-mass T Tauri star TWA 30B (\citealt{2010AJ....140.1486L}; see poster by D.\ Looper). This M4 dwarf exhibits several indicators of an edge-on, actively accreting disk, including strong H~I and alkali emission, forbidden line emission (tracing jet outflows), variable infrared excess, and an overall obscuration of at least 5~mag in the $K$-band relative to its M5 common proper motion companion TWA~30A.  Among the emission lines detected in the optical spectrum of TWA~30B were the 0.873~$\micron$  and 0.985~$\micron$ transitions of [C~I], lines previously detected in planetary nebulae (e.g., \citealt{1983ApJ...268..683J}) and cometary comae (e.g., \citealt{2002ApJ...581..770O}), but not (to our knowledge) in the spectrum of a T Tauri star.  Figure~\ref{fig_twa30} shows a close-up of the FIRE cross-dispersed spectrum of this source obtained on 3 April 2010 (UT); the [C~I] line is clearly present with an equivalent width of $-$12~{\AA}.    In addition, we detected the 1.083~$\micron$ He~I line, broadened to a full width half maximum of $\approx$100~{\kms}.  He~I emission had not been seen in the optical spectrum of this source. The high-excitation [C~I] line, formed in the inner regions of the disk, is an important tracer of gasous disk size when compared to the fine structure, mid-infrared [C~I] lines formed in the colder outer regions of the disk \citep{2009A&A...496..725E}.  He~I emission directly measures accretion, and the unexpected appearance of  this line in the FIRE data suggests episodic accretion or variable obscuration is taking place on TWA~30B.

\subsection{Identification and Kinematics of the Coldest Brown Dwarfs}

\begin{figure}[!ht]
\plotone{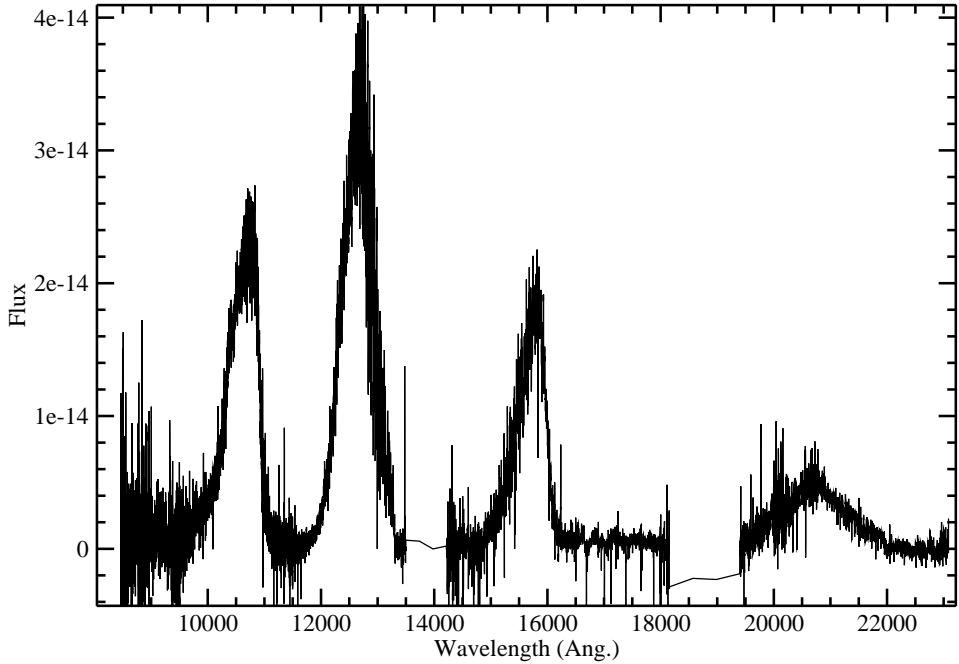}
\caption{Cross-dispersed spectrum of the $J$ = 16.5, {\teff} = 500~K T10 brown dwarf UGPS~J0722-05, based on a preliminary FIRE data reduction pipeline.  Much of the ``noise'' in these data arise from molecular transitions of {\wat} and {\meth} that blanket the near-infrared spectra of cold brown dwarfs (Bochanski et al., in prep.)
\label{fig_ugps}}
\end{figure}

FIRE is ideal for confirmation and kinematic studies of cold, intrinsically faint brown dwarfs.  The prism mode is particularly useful for following up new discoveries anticipated from the Wide-field Infrared Survey Explorer (WISE; e.g., \citealt{2008SPIE.7017E..16L}) and the Visible and Infrared Survey Telescope for Astronomy (VISTA; \citealt{2004Msngr.117...27E}).  During commissioning observations, several promising WISE cold brown dwarf candidates were targeted for observation (see contribution by J.\ D.\ Kirkpatrick).  

The cross-dispersed mode is also useful for extending 3D kinematic studies of ``ultracool'' low mass stars and brown dwarfs (e.g., \citealt{2007ApJ...666.1205Z, 2010AJ....139.1808S}) down to the ``ultracold'' regime (i.e, {\teff} $<$ 800~K).  As an example, Figure~\ref{fig_ugps} displays cross-dispersed observations of the {\teff} $\approx$ 500 K UGPS~J072227.51-054031.2 (hereafter UGPS~J0722-05; \citealt{2010MNRAS.408L..56L}).  With $J$ = 16.5, UGPS~J0722-05 is too faint for high-resolution spectrographs such as Keck/NIRSPEC \citep{2000SPIE.4008.1048M}, but is an easy target with FIRE.  The forest of molecular lines from {\wat} and {\meth} in the near-infrared can be exploited to yield radial velocities accurate to $\lesssim$5~{\kms} through cross-correlation techniques (the native resolution of the data shown in Figure~\ref{fig_ugps} is 50~{\kms}).  We are currently conducting a radial velocity survey of $\sim$70 late-type L and T dwarfs within 20~pc of the Sun, and have obtained multi-epoch spectra for a handful of unresolved L/T binary candidates that are potential radial velocity variables (e.g., \citealt{2008ApJ...681..579B, 2008ApJ...678L.125B, 2010ApJ...710.1142B}).

\subsection{Terrestrial Planet Accretion onto a White Dwarf}

\begin{figure}[!ht]
\plotone{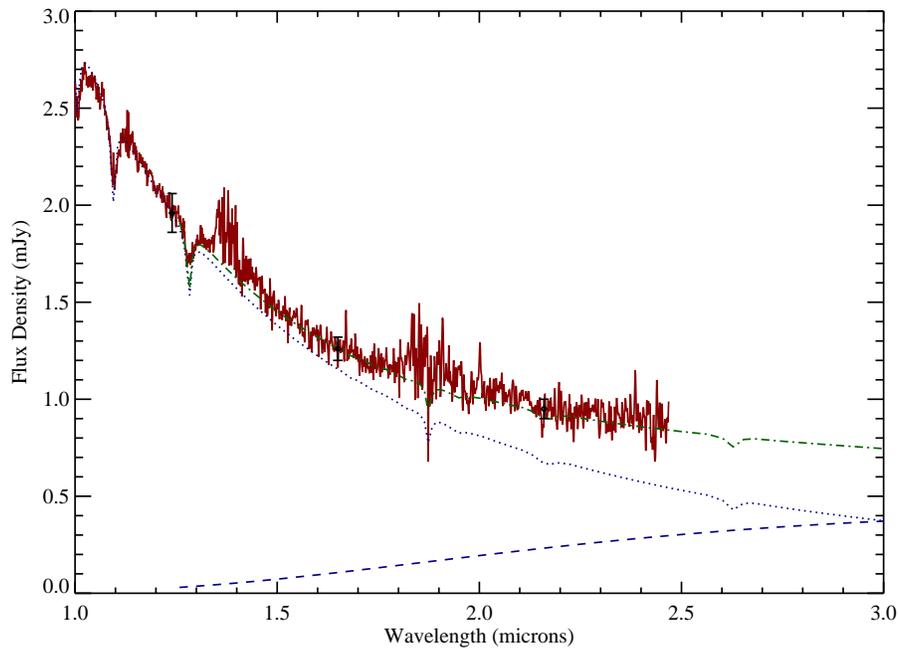}
\caption{Prism-dispersed FIRE spectrum of the DAZ GALEX~1931+01.  The observed 
spectral energy distribution (red line) deviates from the {\teff} = 23,500~K, $\log{g}$ = 8.0~(cgs) atmosphere model (blue dotted line) in a manner consistent with the presence of flat, opaque dust disk (blue dashed line). The green dot-dashed line shows the sum of the atmosphere and disk models (Melis et al., submitted).
\label{fig_galex}}
\end{figure}

The unusually metal-rich spectra of DAZ white dwarfs are attributed to the accretion of metal-rich material, potentially from remnant planetary bodies (e.g., \citealt{2007ApJ...671..872Z, 2010MNRAS.404.2123F}).  Excess infrared emission associated with these stars can also be interpreted as arising from an accretion disk or a low-temperature companion (e.g., \citealt{2005AJ....130.1221D, 2007ApJ...663.1285J}).   A recent case in point is GALEX J193156.8+011745, a heavily polluted DAZ with exceptional abundances of heavy elements and near-infrared excess. \citet{2010MNRAS.404L..40V} attributed these features to the presence of an mid-type L dwarf companion polluting the white dwarf through Roche lobe overflow.  We observed GALEX~1931+01 with FIRE, using both the cross- and prism-dispersed modes.  As shown in Figure~\ref{fig_galex}, these data confirm the near-infrared excess inferred from photometric measurements, but reveal no spectral features indicative of an L or T dwarf companion. Rather, the spectrum indicates the presence of a hot dust disk (i.e., a ``ring of fire''), the remains of tidally-disrupted, terrestrial-like planetary body  (Melis et al., submitted).

\acknowledgements The authors gratefully acknowledge the staff of Las Campanas Observatory for helping to bring FIRE to first light and science operations. AJB acknowledges support from the Chris and Warren Hellman Fellowship. CM acknowledges support by the NSF under award No. AST-1003318.  FIRE's construction was supported by the NSF MRI program, under award number AST-0649190, with additional support from the Curtis Marble Instrumentation Development Fund, the MIT-Kavli Instrument Development Fund, and the MIT Department of Physics.
\bibliography{burgasser_a}

\end{document}